\documentclass[11pt,a4paper]{article}
\usepackage{jheppub}
\usepackage{graphicx}
\usepackage{dcolumn}
\usepackage{bm}

\begin{document}

\title{On Deformed Phase Space  and $\Lambda$.}

\author[a]{M. Sabido}
\author[a]{, S. P\'erez-Pay\'an}
 \author[b]{and E. Mena}
 \affiliation[a]{ Departamento  de F\'{\i}sica de la Universidad de Guanajuato,\\
 A.P. E-143, C.P. 37150, Le\'on, Guanajuato, M\'exico.\\}
 \affiliation[b]{ Centro Universitario de la Ci\'enega, Universidad de Guadalajara\\
Ave. Universidad 1115 Edif. B, C.P. 47820 Ocotl\'an, Jalisco, M\'exico.
\\}

\emailAdd{msabido@fisica.ugto.mx}
\emailAdd{sinuhe@fisica.ugto.mx}
\emailAdd{emena@cuci.udg.mx}

\abstract{In this paper we study the effects of a deformed phase space of an empty (4+1) Kaluza-Klein universe with cosmological constant $\Lambda$. We analyze the effects of the phase space deformations on $\Lambda$, and comment on the possibility that the origin of the cosmological constant in this model is related to the deformation parameter associated to  the 4 dimensional scale factor and the compact extra dimension.
}



\maketitle

\section{Introduction}

The cosmological constant problem has been studied for a long time in different scenarios and has remain as one of the major open problems in modern physics  \cite{polchinski}. It is associated with the cosmic acceleration and in a more general context  with the dark energy problem \cite{dark}. There is a belief that the solution will come from an unconventional approach (i.e. arguments are given that  UV/IR mixing mechanism is needed \cite{polchinski}). 

An unconventional  idea is that of noncommutative spacetime \cite{snyder}. A natural feature of noncommutative quantum field theory is UV/IR mixing \cite{nekrasov}.  The central idea behind  noncommutativity is to express uncertainty in the simultaneous measurement of any pair of conjugate variables, such as position and momentum. Noncommutativity was revived at the beginning of this century \cite{ref3,wess}. This
renewed interest  slowly but steadily  permeated in the realm of gravity, from which  several approaches to noncommutative gravity \cite{ncsdg,ncsdg2} were proposed. The end result of  these formulations  is a highly nonlinear theory,  and finding  solutions to the corresponding noncommutative  field equations is complicated.  

It is argued that noncommutativity can affect the evolution of the universe,  making the study of noncommutative cosmological models an interesting testing ground \cite{wally1}.  An approach to find noncommutative cosmological models, is to derived them from the full noncommutative theory of gravity, this might  seem as a fruitless endeavour, but in order to avoid  difficulties in the study of the possible influence of noncommutativity in relation with the cosmological constant, we will use the  ideas in \cite{ncqc}.  The main ingredients of this proposal are: minisuperspace approach to cosmology  whose variables are the 3-metric components in a finite configuration space. This formalism has the advantage that the inclusion of matter is straightforward. By considering these models one freezes out degrees of freedom and the canonical quantization of these minisuperspace models gives the Wheeler-DeWitt equation (WDW). A general analysis  suggest that  conditions can be found to justify the minisuperspace approach and presume the behavior of the wave function as fundamental \cite{halliwell}. If one considers string theory,  the WDW equation derived from general relativity  corresponds to an $S$-wave approximation \cite{susskind}. Secondly, noncommutative space-time  has the consequence that the fields do not commute. In a more specific manner this is  due to the Moyal product \cite{ref3,wess}. To introduce these elements we take into account the  procedure to generalise usual quantum mechanics to the noncommutative version \cite{gamboa}. Having the WDW equation to describe the quantum evolution of the universe and being the ``coordinates" of these models the fields, it was assumed that the variables  do not commute. Then, an effective noncommutativity was defined in the minisuperspace from which the quantum evolution of the  cosmological model was studied \cite{ncqc}.  This approach is also known as deformed phase space cosmology. In  the last few years there have been several attempts to study the possible effects of 
phase space deformations in the cosmological scenario. In \cite{kallosh} it is argued that there is a possible relation between the 4D cosmological constant and the noncommutative parameter of the compactified space in string theory, also in \cite{darabi} noncommutativity is introduced in a 5 dimensional Kaluza-Klein universe in order to study the hierarchy problem.  In \cite{vakili}, evidence is presented of the relationship between late time acceleration in dilaton cosmology and the deformation parameters. Furthermore, a more direct relation in connection with the cosmological constant problem has been  addressed in \cite{nclambda}, where it is shown that by means of minisuperspace noncommutativity a small cosmological constant arises, and seems to alleviate the discrepancy between the calculated and observed vacuum energy density.

The aim of this  paper is to analyze  the effects of the phase space deformation on the cosmological constant $\Lambda$ and from the resulting deformed phase space model we find a relationship between the cosmological constant $\Lambda$ and the deformation parameters $\theta$ and $\beta$.  

The paper is organised as follows, in section 2, we start with an  empty  (4+1) dimensional Kaluza-Klein universe with cosmological constant and an FRW metric.  In section 3, the noncommutative model is presented,  we introduce  a deformation in the phase space constructed from the minisuperspace variables and their conjugate momenta. Section 4 is devoted for discussion and conslusions.

\section{The Model}

Let us begin introducing the model in a classical scenario which is  an empty $(4+1)$ theory of gravity with cosmological constant $\Lambda$. In this setup the action takes the form
\begin{equation}
I=\int\sqrt{-g}\left(R-\Lambda\right)dtd^3rd\rho,\label{he}
\end{equation} 
where $\left\{t,r^i\right\}$ are the coordinates of the 4-dimensional space time and $\rho$ represents the coordinate of the fifth dimension.
We are interested in Kaluza-Klein cosmology, so an FRW type metric is assumed, which is of the form
\begin{equation}
ds^2=-dt^2+\frac{a^2(t)dr^idr^i}{\left(1+\frac{\kappa r^2}{4}\right)^2}+\phi^2(t)d\rho^2,
\end{equation}  
where $\kappa=0,\pm1$ and $a(t)$, $\phi(t)$ are the scale factors of the universe and the compact dimension respectively. Substituting this metric into the action (\ref{he}) and integrating over the spatial dimensions, we obtain an effective lagrangian that only depends on $(a,\phi)$
\begin{equation}
L=\frac{1}{2}\left(a\phi\dot{a}^2+a^2\dot{a}\phi-\kappa a\phi+\frac{1}{3}\Lambda a^3\phi\right).\label{L1}
\end{equation}
For the purposes of simplicity and calculations, we can rewrite this lagrangian in a more convenient way
\begin{equation}
L=\frac{1}{2}\left[\left(\dot{x}^2-\omega^2 x^2\right)-\left(\dot{y}^2-\omega^2 y^2\right)\right],\label{L2}
\end{equation}
where the new variables were defined as 
\begin{equation}
x=\frac{1}{\sqrt{8}}\left(a^2+a\phi-\frac{3\kappa}{\Lambda}\right),~y=\frac{1}{\sqrt{8}}\left(a^2-a\phi-\frac{3\kappa}{\Lambda}\right),\label{3}
\end{equation}
and $\omega^2=-\frac{2\Lambda}{3}$.
The Hamiltonian for the model is calculated from the usual relation from classical mechanics $H=p_i\dot{q}_i-L(q_i,p_i)$, where $q_i$ are the generalized coordinates and $p_i$ the conjugate momenta. Therefore the Hamiltonian reads  
\begin{equation}
H=\left[\left(p_{x}^2+\omega^2 x^2\right)-\left(p_{y}^2+\omega^2 y^2\right)\right],\label{HC}
\end{equation}
which describes an isotropic oscillator-ghost-oscillator system. By following the canonical formalism, from (\ref{HC}) we can construct the WDW equation, and get  the corresponding quantum cosmology for the model at hand. This is achieved by making the usual identifications $p_x=-i\partial/\partial{x}$ and $p_y=-i\partial/\partial{y}$,
\begin{equation}
\left[\left(-\frac{\partial^2}{\partial^2y}-\omega^2 y^2\right)-\left(-\frac{\partial^2}{\partial^2x}-\omega^2 x^2\right)\right]\Psi(x,y)=0.\label{ham}
\end{equation}
This equation can gives the quantum description of the  cosmological model and the information about the quantum behaviour would be encoded in the wave function $\Psi(x,y)$.

\section{Noncommutative Model}

As is well known, there are different approaches to include noncommutativity to physical  theories. In particular, to study noncommutative cosmology, there exist a well explored path to study noncommutativity in a cosmological setting  \cite{ncqc}.  In this set up the noncommutativity is realized  in the minisuperspace variables.

In canonical quantum cosmology, after canonical quantization, one formally obtains the Wheeler-deWitt equation. This is a Klein-Gordon type equation which describes the quantum behaviour of the universe. An alternative approach  to study quantum mechanical effects, is to introduce deformations to the phase space of the system. The approach is an equivalent path to quantization and is part of a complete and consistent type of quantization known as deformation quantization \cite{hugo}. Our interest is in cosmology and  these models are constructed in the minisuperspace, following the previous discussion we can assume the studying cosmological models in deformed phase could be interpreted as studying quantum effects to cosmological solutions \cite{darabi}.
In the deformed phase space approach, the deformation is introduced by the Moyal brackets $\{f,g\}_{\alpha}=f\star_{\alpha}g-g\star_{\alpha}f$, were the product between functions is replaced by  the Moyal product
$
(f\star{g})(x)=\exp{\left[\frac{1}{2}\alpha^{ab}\partial_{a}^{(1)}\partial_{b}^{(2)}\right]}f(x_{1})g(x_2)\vert_{x_1=x_2=x}
$
such that
\begin{eqnarray}
\alpha =
\left( {\begin{array}{cc}
 \theta_{ij} & \delta_{ij}+\sigma_{ij}  \\
- \delta_{ij}-\sigma_{ij} & \beta_{ij}  \\
 \end{array} } \right),
\end{eqnarray}
where the $2\times 2$ matrices $\theta_{ij}$ and $\beta_{ij}$ are assumed to be antisymmetric and represent the noncommutativity in the coordinates and momenta respectively. The resulting  $\alpha$ deformed algebra for the phase space variables is
\begin{equation}
\{x_i,x_j\}_{\alpha}=\theta_{ij}, \;\{x_i,p_j\}_{\alpha}=\delta_{ij}+\sigma_{ij},\; \{p_i,p_j\}_{\alpha}=\beta_{ij}.\label{alg}
\end{equation}

In this paper we consider particular expressions for the deformations, namely $\theta_{ij}=-\theta\epsilon_{ij}$ and $\beta_{ij}=\beta\epsilon_{ij}$.

Let us consider and alternative to derive a similar algebra to Eq.(\ref{alg}). The resulting algebra will be the  same, but the Poisson brackets are different in the two algebras. For Eq.(\ref{alg})  the brackets are the $\alpha$ deformed ones and are related to the  Moyal product, for the other algebra the brackets are the usual Poisson brackets.

Making the following transformation on the classical phase space variables $\{x,y,P_x,P_y\}$
\begin{eqnarray}
\hat{x}=x+\frac{\theta}{2}P_{y}, \qquad \hat{y}=y-\frac{\theta}{2}P_{x},\nonumber \\
\hat{P}_{x}=P_{x}-\frac{\beta}{2}y, \qquad \hat{P}_{y}=P_{y}+\frac{\beta}{2}x, \label{nctrans}
\end{eqnarray}
the algebra reads
\begin{equation}
\{\hat{y},\hat{x}\}=\theta,\; \{\hat{x},\hat{P}_{x}\}=\{\hat{y},\hat{P}_{y}\}=1+\sigma,\; \{\hat{P}_y,\hat{P}_x\}=\beta,\label{dpa}
\end{equation}
where $\sigma=\theta\beta/4$. For the rest of this paper we will use the algebra in Eq.(\ref{dpa}). Now that we have constructed the modified phase space, we start with a Hamiltonian which is formally analogous to Eq.(\ref{HC}) but constructed with the variables $(\hat{x},\hat{y},\hat{P}_x,\hat{P}_y)$ that obey the modified algebra (\ref{dpa}). After using the change of variables (\ref{nctrans}) the deformed Hamiltonian in terms of the commutative variables is found to be
\begin{eqnarray}
 H&=&N \left( \frac{1}{2} \hat{P}_x^2 + \frac{\omega^2}{2} \hat{x}^2\right) -N \left( \frac{1}{2} \hat{P}_y^2 + \frac{\omega^2}{2} \hat{y}^2 \right)\\
&=& \frac{1}{2}\left[\left( p_{x}^2-p_{y}^2\right )-\omega_{1}^2(x p_{y}+ yp_{x}) + \omega_{2}^2(x^2- y^2)\right] . \label{HNC}\nonumber
\end{eqnarray}
In order to obtain the previous expression we have used the following definitions
\begin{equation}
\omega_{1}^2=\frac{\beta-\omega^2\theta}{1-\frac{\omega^2\theta^2}{4}}, \quad\omega_{2}^2=\frac{\omega^2-\frac{\beta^2}{4}}{1-\frac{\omega^2\theta^2}{4}}.\label{omegas}
\end{equation}
We can construct a bidimensional vector potential ${A}_{x}=\frac{\omega_1^2}{2}{y}$, ${A}_{y}=-\frac{\omega_1^2}{2}{x}$,  from $\bf{B=\nabla\times{A}}$ and the vector potential $\bf A$ we find a magnetic field $B=-\omega_1^2$ hence the vector potential can be rewritten  as $A_{{x}}=-\frac{B}{2}{y}$ and  $A_{{y}}=\frac{B}{2}{x}$. On the other hand, we already know from (\ref{dpa}) that $ [\hat{P}_y,\hat{P}_x]=\beta$ and if we set $\theta=0$ in the above equation for $B$ we can conclude that the deformation of the momentum plays a role analogous to a magnetic field. This result allow us to write the effects of the noncommutative deformation as minimal coupling on the Hamiltonian, ${H}=\frac{1}{2}[(p_{x}-{A}_{x})^2+\omega_3^{ 2}{x}^2]-\frac{1}{2}[(p_{y}-{A}_{y})^2+\omega_3^{2}{y}^2]$, where $\omega^{2}_{3}=\frac{\omega_{1}^4}{4}+\omega_{2}^2$.  We can write the hamiltonian in terms of the magnetic B-field as
\begin{equation}
{H}=\frac{1}{2}\left(p_{x}+\frac{B}{2}y\right)^2-\frac{1}{2}\left(p_{y}-\frac{B}{2}x\right)^2
+\frac{1}{2}\left(\omega_2^{ 2}+\frac{B^2}{4}\right)\left({x}^2-{y}^2\right).\label{14} 
\end{equation}
As this theory is now described by the commutative variables,  we can interpret the effects of the  deformation as the presecence of the B field. 
After a closer inspection of the equation, it is convenient to  rewrite (\ref{14}) in a much simpler and suggestive form
\begin{equation}
H=\left({p}_{x}^2-{p}_{y}^2\right)+\left(\omega_3^{2}-\frac{B^2}{4}\right)\left({x}^2-{y}^2\right)
+B\left({y}{p}_{x}+{x}{p}_{y}\right),\label{15}
\end{equation}
which is a two dimensional anisotropic ghost-oscillator \cite{darabi}. 

\section{Discussion}
We have done a deformation of the phase space of the theory,  this gives two new fundamental constants $\theta$ and $\beta$. As we have stablished,  the frequency  $\omega$ is related to the cosmological constant, therefore we are interested in  the effective frequency of the deformed model. In order to achieve this we must determine the effects of the  last term in Eq.(\ref{15})  ( ``B-term''). To understand the effects of the B-term on the oscillator frequency we proceed as in \cite{darabi}.  First we compare the ghost oscillator with the two dimensional harmonic oscillator endowed with the vector potential $A=(-\frac{B}{2}{y},\frac{B}{2}{x})$ and oscillator frequency $\omega$. The resulting Hamiltonian is
\begin{equation}
H=\left({p}_{x}^2+{p}_{y}^2\right)+\left(\omega^{ 2}+\frac{B^2}{4}\right)\left({x}^2+{y}^2\right) 
+B\left({y}{p}_{x}-{x}{p}_{y}\right).\label{16}
\end{equation}
For the two dimensional oscillator system, the quantity $\sqrt{\omega^{ 2}+\frac{B^2}{4}}$ is the oscillator frequency and the B-term represents the magnetic potential energy. This B-term has an independent role and does not modify the oscillator frequency. From a comparison between the two systems,  we can see  a correspondence between the terms  in Eq.(\ref{15}) and Eq.(\ref{16}). 
The first two terms in Eq.(\ref{16}) correspond to the first two terms in Eq.(\ref{15}), and the B-term in the oscillator system corresponds to the B-term of the ghost oscillator. Then we conclude that the corresponding terms play the same role, and therefore the B-term in the ghost oscillator plays an independent role (as in the two dimensional oscillator) and does not modify the frequency.

With this in mind, we can define the deformed ghost oscillator frequency $\tilde{\omega}$, 
\begin{equation}
\tilde{\omega}^2=\omega_3^{2}-\frac{B^2}{4}=\frac{4\omega^2-\beta^2}{4-\omega^2\theta^2}, \qquad \omega^2\theta^2\le1, \label{omegatilde}
\end{equation}
in section II, $\omega$ was defined from the cosmological constant, as in \cite{darabi} we assume that for the deformed model  $\tilde{\omega}$ is obtained by a new effective cosmological constant  $\tilde\Lambda_{eff}=-\frac{3}{2}\tilde{\omega}^2$. With these definitions we get 
\begin{equation}
\tilde\Lambda_{eff}=\frac{4\Lambda+\frac{3}{2}\beta^2}{4-\frac{2}{3}\theta^2\mid\Lambda\mid}.\label{eq18}
\end{equation} 
The case $\beta=0$ reduces to the noncommutative minisuperspace model, the effective cosmological constant 
is modified by the noncommutative parameter $\theta$, but this parameter can not replace the cosmological constant. In \cite{darabi} it was used to present a solution to the Hierarchy problem. The author assumed  the electroweak scale $M_{EW}$ to be the natural cutoff in the commutative model and the Planck scale $M_p$  as the cutoff in the noncommutative model. Then by using Eq.(\ref{eq18}) with vanishing $\beta$, the two scales are related by the parameter $\theta$, explaining the Hierarchy problem in the context of the Wheeler-DeWitt equation.

Now we  turn our attention to the case where there is no deformation on the coordinates. Taking  the noncommutative parameter $\theta=0$ we have that the frequency and the effective cosmological constant are given by
\begin{equation}
\tilde{\omega}^2=\omega^2-\frac{\beta^2}{4},\qquad \textrm {and} \qquad
\tilde{\Lambda}_{eff}=\Lambda+\frac{3\beta^2}{8}.
\end{equation}
From the last equation we get the most interesting result of this paper.  We can see that the deformation parameter $\beta$ and the cosmological constant  $\Lambda$ compete to give the effective cosmological constant $\tilde{\Lambda}_{eff}$. Therefore, if we start with a commutative anti-de Sitter model, because of the deformation parameter $\beta$ we could have a de-Sitter.
If we consider the case of a flat universe with  a vanishing $\Lambda$ we see that $\tilde{\Lambda}_{eff}=\frac{3\beta^2}{8}$.  We get and effective the cosmological constant  from the parameter $\beta$.
As already argued, the presence of the deformation parameters $\theta$ and $\beta$ gives an unconventional origin to the cosmological constant. In this approach the cosmological constant is given strongly related to the parameter $\beta$, which is considered a new fundamental constant. In this case, one can argue that the origin of $\Lambda$  is related to the noncommutative deformation between the canonical momenta associated to the 4 dimensional scale factor and the canonical momenta associated to the compact extra dimension. Recently, some evidence on the possibility that the effects of the phase space deformation could be related to the late time acceleration of the universe as well as to the cosmological constant was presented \cite{vakili}.

Interestingly, in the particular case of $\beta=\omega^2\theta$ we find  that frequency reduces to $\tilde{\omega}^{2}=\omega^2$ and the magnetic potential energy vanishes as well as the effective magnetic field. Then  we have that $\tilde{\Lambda}_{eff}={\Lambda}$ and in  this case even as we have done a deformation on the minisuperspace of the theory, the effects cancel out and the resulting theory behaves as in the commutative theory.

\section*{Acknowledgments}
M.S.  supported by DAIP grant 125/11 and  CONACYT grants 62253, 135023. S. P. P. is supported by CONACyT graduate grant. E.M. is partially supported by  PROMEP grants 103.5/10/6209. This work is part of the PROMEP research network  ``Gravitaci\'on y F\'isica Matem\'atica".

\end{document}